\documentclass[aps,pre,twocolumn,groupedaddress,showpacs,amssymb,amsmath]{revtex4-1}
\usepackage{prettyref}
\usepackage{graphicx}
\usepackage{verbatim}   
\usepackage{color}      
\usepackage{subfigure}  
\usepackage{hyperref}
\begin{document}

\preprint{This line only printed with preprint option}

\title{Work distribution under continuous
quantum histories}

\author{Huanan Li}

\email{g0900726@nus.edu.sg}

\affiliation{Department of Physics and Center for Computational Science and Engineering,
National University of Singapore, Singapore 117542, Republic of Singapore }

\author{Jian-Sheng Wang}

\affiliation{Department of Physics and Center for Computational Science and Engineering,
National University of Singapore, Singapore 117542, Republic of Singapore }

\date{23 April 2013}
\begin{abstract}
We present a general scheme to obtain work distribution in closed
systems under continuous quantum histories of corresponding ``power''
operator. The scheme is tested by analytically calculating the quantum
work distribution for a prototype model of a center-shifted one-dimensional
quantum harmonic oscillator. Furthermore, its relationship to the
path integral formalism with the Wick rotation is immediately noticed.
We find that, based on the continuous measurement, generally
Jarzynski equality in quantum regime is not valid, though it is recovered
in high-temperature limit. Also we briefly explore the extension to
open systems, in which the work done and heat transfer
can be calculated in the same set of quantum histories. Our results
for this model can be directly verified by single-molecule
force spectroscopy.
\end{abstract}


\maketitle


The discovery of Jarzynski's equality (JE) \cite{Jarzynski1997} inspires
the intense study for closed systems far from equilibrium over recent
years \cite{Campisi2011,Seifert2012}. The precise mathematical expression
of JE in closed systems irrespective of classical or quantum nature
is clear, which reads

\begin{alignat}{1}
-\frac{1}{\beta}\ln\left\langle e^{-\beta W}\right\rangle = & F(\lambda_{t_{f}})-F(\lambda_{t_{0}}),\label{eq:the JE}
\end{alignat}
although some authors also gave an interesting quantum-mechanical
analog of the JE by introducing a new reference free energy \cite{Ngo2012}.
The foremost condition for JE to be valid requires that in the beginning
of the drive $\lambda_{t}$ the system is in thermal equilibrium with
the bath, whose inverse temperature is $\beta\equiv\left(k_{B}T\right)^{-1}$.
Here $F(\lambda_{t_{0}})$ and $F(\lambda_{t_{f}})$ are the equilibrium-state
free energy of the system corresponding to the system Hamiltonian
at the initial time $t_{0}$ and the finite time $t_{f}$ of the drive
protocol, respectively. However, it should be noticed that the system
is in general not in equilibrium at the end of the drive process.
Now, all the subtleties of deriving and applying the JE are to properly
define work done $W$ and obtain the work distribution $\mathrm{Pr}\left(W\right)$ in
order for the calculation of $\left\langle e^{-\beta W}\right\rangle \equiv\int\mathrm{d}W\mathrm{Pr}\left(W\right)e^{-\beta W}$.
The exploration of the work done $W$ on nonequilibrium systems mainly
follows two lines.

Classically, according to Jarzynski, work done $W$ is well defined,
which can be evaluated as $W=\int_{t_{0}}^{t_{f}}\mathrm{d}t\left[\partial H\left(\lambda_{t}\right)/\partial t\right]$
\cite{Jarzynski2007}. Then a derivation based on Hamiltonian evolution
for thermodynamically isolated systems is feasible \cite{Jarzynski2008},
while the most derivations of the JE for open stochastic classical
systems weakly connected to a heat bath relied on the principle of
detailed balance \cite{Crooks1998}.

Quantum-mechanically, for closed systems perhaps the most successful
operational definition of quantum work is based on the two-time measurement
scheme \cite{Talkner2007}, by which the JE can be proved in a few
lines. But we consider that consistency conditions need to be imposed
on this scheme for the probability we calculate to really make sense
\cite{Li2012}. We emphasize here that the definition of the two-time
measurement protocol for quantum work is operational and expedient,
since what one studies is actually the energy difference, which is
equal to the work done only for closed systems and can not be extended
to open systems due to the presence of heat transfer.

Suppose the evolution of an open quantum system is governed by the
explicitly time-dependent Hamiltonian $\hat{H}\left(t\right)$ and
its state is described by the reduced density matrix $\hat{\rho}\left(t\right)$.
We know that the internal energy of this open system is just $\left\langle \hat{H}\left(t\right)\right\rangle =\mathrm{Tr}\left(\hat{\rho}\left(t\right)\hat{H}\left(t\right)\right)$.
And due to the interaction with the environment, the reduced density
matrix $\hat{\rho}\left(t\right)$ satisfies a master equation $\mathrm{d}\hat{\rho}\left(t\right)/\mathrm{d}t=\frac{1}{i\hbar}\left[\hat{H}\left(t\right),\hat{\rho}\left(t\right)\right]+\mathcal{L}(\hat{\rho})$.
Thus after the differential of $\left\langle \hat{H}\left(t\right)\right\rangle $,
we have the quantum version of the first law of thermodynamics, $\mathrm{d}\left\langle \hat{H}\left(t\right)\right\rangle =\mathrm{Tr}\left(\hat{\rho}\left(t\right)\frac{\partial\hat{H}\left(t\right)}{\partial t}\right)\mathrm{d}t+\mathrm{Tr}\left(\mathcal{L}(\hat{\rho})\hat{H}\left(t\right)\right)\mathrm{d}t$,
in which the first term $\mathrm{Tr}\left(\hat{\rho}\left(t\right)\frac{\partial\hat{H}\left(t\right)}{\partial t}\right)\mathrm{d}t$
can be identified as the work done during the infinitesimal time step, and the second term as heat absorbed.
Once again we notice the key role played by the expression $\frac{\partial\hat{H}\left(t\right)}{\partial t}$
in defining work. For a direct application of this idea, one can resort
to Ref. \cite{Solinas2013}.

In this work, we introduce a different scheme for quantum work based
on continuous quantum histories. The continuous quantum histories
are obtained by continuous measurements of the power operator $\frac{\partial\hat{H}\left(t\right)}{\partial t}$
using von Neumann's prescription, which collapses the wave function
according to the result of measured variable directly (strong measurement).
To our knowledge, a similar idea was first employed by Chernyak and
Mukamel \cite{Chernyak2004} to study the effect of quantum collapse
on the distribution of work. They claimed: `` $\left[\ldots\right]$
more dramatic effects are obtained when using von Neumann's prescription
$\left[\ldots\right]$ We then predict the breakdown of the Gaussian
profile of work and the appearance of long algebraic tails which are
sensitive to the measurement error bar.'' In addition, in their subsequent
paper \cite{Chernyak2006} they thought that ``the measurement of
an observable with a continuous spectrum requires the introduction
of a finite error bar $\varepsilon$. A precise measurement ($\varepsilon\rightarrow0$)
is not properly defined in the quantum case.'' However, our approach
implicitly involves the use of Wick rotation, giving a different prediction
of the distribution of work in the case of strong measurements. In
the following, we first introduce the theory to calculate the
quantum work distribution. Then we apply the approach to a prototype
model of a center-shifted one-dimensional quantum harmonic oscillator,
where the treatment of the Wick rotation will be justified. Finally,
we discuss the consequences of this definition of quantum work and
propose a possible extension to open systems.


$Theory$\rule[0.5mm]{0.05\columnwidth}{1pt}Let us consider a general Hamiltonian $\hat{H}\left(t\right)$, which
is explicitly time-dependent due to a control parameter $\lambda\left(t\right)$.
Then we can define the relevant power operator to work done as $\hat{A}\left(t\right)\equiv\partial\hat{H}\left(t\right)/\partial t=\dot{\lambda}\partial\hat{H}\left(t\right)/\partial\lambda$,
where an over-dot denotes the derivative with respect to time. Since definitely
the power operator $\hat{A}\left(t\right)$ is hermitian, we can find
out the complete set for the Hilbert space consisting of its instantaneous
eigenstates $\left|a\left(t\right)\right\rangle $ satisfying $\hat{A}\left(t\right)\left|a\left(t\right)\right\rangle =a\left(t\right)\left|a\left(t\right)\right\rangle $
and $\left\langle a\left(t\right)\right.\left|a'\left(t\right)\right\rangle =\delta_{aa'}$.
Here $a\left(t\right)$ is the corresponding instantaneous eigenvalues
at time $t$, which we assume are discrete without loss of generality.

As we know, an event in quantum mechanics is specified by a projector.
For example, the projector $\hat{P}_{j}^{a_{j}}\equiv\left|a(t_{j})\right\rangle \left\langle a(t_{j})\right|$
says that at time $t_{j}$ we measure the power operator $\hat{A}\left(t_{j}\right)$
and obtain the eigenvalue to be $a(t_{j})$. Furthermore, repeated
quantum measurements at different times give us a realization of quantum
histories, say,
\begin{align}
Y= & \hat{P}_{0}^{a_{0}}\odot\hat{P}_{1}^{a_{1}}\odot\cdots\odot\hat{P}_{f}^{a_{f}},\label{eq:quantum history}
\end{align}
which means that at the initial time $t_{0}$ the measurement result
of the power operator $\hat{A}\left(t_{0}\right)$ is $a(t_{0})$
with the wave function collapsing to corresponding eigenstate $\left|a\left(t_{0}\right)\right\rangle $,
then the system evolves to the next time $t_{1}$, when the second
measurement is performed, after that we repeat similar intermediate
steps according to the self-evident notations until the final time
$t_{f}$. Where the time duration from the initial time $t_{0}$ to
the final time $t_{f}$ is discretized uniformly so that $t_{j}=t_{0}+j\cdot\Delta t$
with $\Delta t=\left(t_{f}-t_{0}\right)/f$, and $\odot$ is a variant
of the tensor product symbol $\otimes$, emphasizing that the factors
in the quantum history refer to different times \cite{Griffiths2002}.

By convention, the hermitian conjugate of the chain operator associated
with the quantum history shown in Eq. \eqref{eq:quantum history}
is defined as
\begin{align}
K^{\dagger}(Y)=\: & \!\!\hat{P}_{0}^{a_{0}}U(t_{0},t_{1})\hat{P}_{1}^{a_{1}}U(t_{1},t_{2})\cdots U(t_{f-1},t_{f})\hat{P}_{f}^{a_{f}},
\end{align}
obtained by replacing $\odot$s with proper time-evolution operators.
Then it is straightforward but nontrivial to show that the joint probability
distribution of the quantum event $Y$ is
\begin{align}
\mathrm{Pr}(Y)= & \mathrm{Tr}[\rho^{ini}K^{\dagger}(Y)K(Y)],
\end{align}
where the initial density matrix $\rho^{ini}=e^{-\beta\hat{H}(t_{0})}/\mathrm{Tr}(e^{-\hat{H}(t_{0})})$,
after considering the foremost application condition of JE. Naturally,
the work distribution under the quantum histories takes the form:
\begin{align}
\mathrm{Pr}\left(W\right)= & \sum_{a_{0}}\sum_{a_{1}}\cdots\sum_{a_{f}}\delta\left(W-\sum_{j=0}^{f-1}a_{j}\Delta t\right)\mathrm{Pr}(Y)\nonumber \\
= & \int_{-\infty}^{\infty}\frac{\mathrm{d}y}{2\pi}e^{iyW}\sum_{a_{0}}\sum_{a_{1}}\cdots\sum_{a_{f}}e^{-iy\sum_{j=0}^{f-1}a_{j}\Delta t}\mathrm{Pr}(Y)\nonumber \\
\equiv & \int_{-\infty}^{\infty}\frac{\mathrm{d}y}{2\pi}e^{iyW}F\left(y\right),\label{eq:work distribution}
\end{align}
where the integral expression for Dirac $\delta$ function has been
used and $F\left(y\right)$ is just the characteristic function of
$\mathrm{Pr}\left(W\right)$.

The procedure for calculation is a little tricky and subtle: firstly,
when calculating the multiple summations with respect to $a_{j},\: j=0,1,\ldots,f$
or multiple integrals in case of continuous spectrum of power operator,
$\mathit{i.e.}$, $F\left(y\right)$, we introduce the Wick rotation,
i.e., $t\rightarrow i\tau$ and $\Delta t\rightarrow i\Delta\tau$
$\left(\Delta\tau>0\right)$ and let $\Delta\tau\rightarrow0^{+}$
corresponding to the continuous quantum histories (measurement); secondly,
when calculating the final integral with respect to $y$, we transform
the time back so that $\tau=-it$; Lastly, the pre-coefficient of
proportionality is determined by the normalization condition
for $\mathrm{Pr}\left(W\right)$, $\mathit{i.e.}$, $\int\mathrm{d}W\mathrm{Pr}\left(W\right)=1$.


$Application\; and \; justification$\rule[0.5mm]{0.05\columnwidth}{1pt}While it is straightforward to use a collection of harmonic oscillators
to test the theory, for the results to be physically clear we simply
choose the model a center-shifted one-dimensional quantum harmonic
oscillator, which is relevant to the experiment of a driven single
molecule \cite{Hummer2005}. Its Hamiltonian reads

\begin{align}
\hat{H}(t)= & \frac{\hat{p}^{2}}{2m}+\frac{1}{2}m\omega^{2}\left[\hat{x}-\lambda\left(t\right)\right]^{2},
\end{align}
and for convenience, we set $\lambda\left(t_{0}\right)=0$. Then we
can directly obtain the power operator $\hat{A}\left(t\right)=-\dot{\lambda}\left(t\right)m\omega^{2}\left[\hat{x}-\lambda\left(t\right)\right]$,
which satisfies
\begin{align}
\hat{A}\left(t\right)\left|x\right\rangle = & -\dot{\lambda}\left(t\right)m\omega^{2}\left[x-\lambda\left(t\right)\right]\left|x\right\rangle .
\end{align}
Here $\left|x\right\rangle $ is just the eigenstate of the position
operator $\hat{x}$. In this example, the joint probability distribution
of the quantum event $Y=\hat{P}_{0}^{x_{0}}\odot\hat{P}_{1}^{x_{1}}\odot\cdots\odot\hat{P}_{f}^{x_{f}}$
is
\begin{multline}
\mathrm{Pr}(Y)=\left\langle x_{0}\right|\rho^{ini}\left|x_{0}\right\rangle \left|\left\langle x_{0}\right|e^{\frac{i}{\hbar}H_{1}\Delta t}\left|x_{1}\right\rangle \right|^{2}\\
\left|\left\langle x_{1}\right|e^{\frac{i}{\hbar}H_{2}\Delta t}\left|x_{2}\right\rangle \right|^{2}\cdots\left|\left\langle x_{f-1}\right|e^{\frac{i}{\hbar}H_{f}\Delta t}\left|x_{f}\right\rangle \right|^{2}
\end{multline}
with $H_{j}=\hat{H}\left(t_{j}\right)$. The calculation of the transition
probability $\left|\left\langle x_{j}\right|e^{\frac{i}{\hbar}H_{j+1}\Delta t}\left|x_{j+1}\right\rangle \right|^{2}$
$j=0,1,\ldots f-1$ requires special attention. For this model, assuming
$\Delta t\rightarrow0$ we can easily get
\begin{align}
 & \left\langle x_{j}\right|e^{\frac{i}{\hbar}H_{j+1}\Delta t}\left|x_{j+1}\right\rangle \nonumber \\
= & \sqrt{\frac{im}{2\pi\hbar\Delta t}}e^{-\frac{i}{\hbar}\Delta t\left[\frac{m}{2}\left(\frac{x_{j}-x_{j+1}}{\Delta t}\right)^{2}-\frac{m\omega^{2}}{2}\left(x_{j+1}-\lambda_{j+1}\right)^{2}\right]}.
\end{align}
 Where $\lambda_{j+1}=\lambda(t_{j+1})$. If we naively use this expression
to calculate $\left|\left\langle x_{j}\right|e^{\frac{i}{\hbar}H_{j+1}\Delta t}\left|x_{j+1}\right\rangle \right|^{2}$,
some troubles will immediately come. In that circumstances, the probability
$\left|\left\langle x_{j}\right|e^{\frac{i}{\hbar}H_{j+1}\Delta t}\left|x_{j+1}\right\rangle \right|^{2}=\frac{m}{2\pi\hbar\Delta t}$
does not depend on the two measurement results $x_{j}$ and $x_{j+1}$
separated by infinitesimal time difference $\Delta t$, which is physically
unreasonable. In other words, roughly speaking this result does not
satisfy the mathematical requirement $\left|\left\langle x_{j}\right|e^{\frac{i}{\hbar}H_{j+1}\Delta t}\left|x_{j+1}\right\rangle \right|^{2}=\delta\left(x_{j}-x_{j+1}\right)^{2}$
when $\Delta t\rightarrow0$. But one should not take it too seriously,
the square of the Dirac distribution being meaningless. Furthermore
one can convince himself that this argument is generally true irrespective
of the specific form of the Hamiltonian. Therefore, as the theory
part said, we introduce the Wick rotation to conquer this trouble
and essentially this procedure is responsible for the multiple integrals
in the characteristic function $F\left(y\right)$
to be convergent.

After introducing Wick rotation, $\mathrm{i.e.}$, setting $t_{j}\rightarrow i\tau_{j}$
and $\Delta t\rightarrow i\Delta\tau$ $\left(\Delta\tau\rightarrow0^{+}\right)$,
the multiple integrals $F\left(y\right)$ in the case of continuous
spectrum of power operator becomes
\begin{multline}
F\left(y\right)\propto\int\mathrm{d}x_{0}\mathrm{d}x_{1}\cdots\mathrm{d}x_{f-1}\exp\bigg[-y\Delta\tau\sum_{j=0}^{f-1}\dot{\lambda}_{j}\left(x_{j}-\lambda_{j}\right)\\
-x_{0}^{2}\tanh\frac{\beta}{2}-\sum_{j=0}^{f-2}\frac{1}{\Delta\tau}\left(x_{j}-x_{j+1}\right)^{2}-\sum_{j=0}^{f-2}\Delta\tau\left(x_{j+1}-\lambda_{j+1}\right)^{2}\bigg]\\
\propto\exp\bigg[\frac{y^{2}\left(\sum_{j=0}^{f-1}\dot{\lambda}_{j}\Delta\tau\right)^{2}+4y\tanh\frac{\beta}{2}\sum_{j=0}^{f-1}\lambda_{j}\dot{\lambda}_{j}\Delta\tau}{4\tanh\frac{\beta}{2}}\bigg],
\end{multline}
where $\dot{\lambda}_{j}=\dot{\lambda}(t_{j})$ and we have adopted
the natural units that $\hbar=m=\omega=1$ and used the result $\left\langle x_{0}\right|\rho^{ini}\left|x_{0}\right\rangle \propto e^{-x_{0}^{2}\tanh\frac{\beta}{2}}$
for the equilibrium state of a one-dimensional quantum harmonic oscillator.
There are two points which need to be clarified: first, the integral
with respect to $x_{f}$ has been performed in advance since the prefactor
$e^{-iy\sum_{j=0}^{f-1}a_{j}\Delta t}$ before $\mathrm{Pr}(Y)$ generally
does not depend on $a_{f}$; second, the proportionality constant
may be divergent when $\Delta\tau\rightarrow0^{+}$ since the eigenstate
of the power operator $\left|x\right\rangle $ can not be properly
normalized. However what we are concerned is the relative probability
distribution and the final proportionality constant will be fixed
by the normalization condition of $\mathrm{Pr}\left(W\right)$.

Now it is fairly easy to calculate $\mathrm{Pr}\left(W\right)$ under
continuous quantum histories. But before performing the final integral
with respect to $y$, we have to transform the time back, i.e., $\tau=-it$.
Thus, employing Eq. \eqref{eq:work distribution} and writing the
units back carefully, the work distribution for this prototype model
after normalization is given as

\begin{align}
\mathrm{Pr}\left(W\right)= & \sqrt{\frac{\tanh\frac{\beta}{2}\hbar\omega}{\pi\overline{\dot{\lambda}}^{2}m\hbar\omega^{3}}}\exp\bigg[-\frac{\left(W-\overline{\lambda\dot{\lambda}}m\omega^{2}\right)^{2}\tanh\frac{\beta}{2}\hbar\omega}{\overline{\dot{\lambda}}^{2}m\hbar\omega^{3}}\bigg]\label{eq:work distribution Eg}
\end{align}
with $\lambda\left(t_{0}\right)=0$, $\overline{\dot{\lambda}}\equiv\int_{t_{0}}^{t_{f}}\mathrm{d\lambda}$
and $\overline{\lambda\dot{\lambda}}\equiv\int_{t_{0}}^{t_{f}}\lambda d\lambda$
. Notice that strictly speaking $\overline{\dot{\lambda}}$ and $\overline{\lambda\dot{\lambda}}$
are both Riemann\textendash{}Stieltjes integrals and if the profile
of the control parameter $\lambda\left(t\right)$ is smooth enough
with respect to time $t$, this work distribution does not rely on
the driven process by the control parameter due to the nature of this
simple model. On the contrary, Eq. \eqref{eq:work distribution Eg}
can be also applied to a stepwise pulling protocol for $\lambda\left(t\right)$
and the result for the limiting case of infinite pulling steps in
a finite time duration can be considered the same as Ref. \cite{Ngo2012},
in which the authors considered that the system relaxes to the thermal
equilibrium related to the updated Hamiltonian in every pulling step
due to the weak coupling to the heat bath.


$Consequence\; and\; extention$\rule[0.5mm]{0.05\columnwidth}{1pt}First, let us consider an interesting situation, in which the control
parameter $\lambda\left(t\right)$ vanishes identically. Then according
to Eq. \eqref{eq:work distribution Eg}, we know that $\mathrm{Pr}\left(W\right)$
approaches $\delta\left(W\right)$, which means that the operation
of continuous measurement does not do any work to the system. In other
words, the work done is solely due to the drive $\lambda\left(t\right)$
under continuous quantum histories. This fact is reasonable which
reflects the quantum Zeno effect, $i.e.,$ continuous measurement
of the system will keep the system fixed at the starting eigenstate
of the observable corresponding to the initial measurement result
\cite{Misra1977}.

Perhaps the most important consequence of the quantum work under continuous
quantum histories is that the JE generally fails. To illustrate this
point, we calculate the left-hand side of the Eq. \eqref{eq:the JE}
using the work distribution Eq. \eqref{eq:work distribution Eg}.
And after assuming smoothness of the profile of the control parameter,
we obtain
\begin{align}
-\frac{1}{\beta}\ln\left\langle e^{-\beta W}\right\rangle = & \lambda_{f}^{2}m\omega^{2}\left(\frac{1}{2}-\frac{\beta\hbar\omega}{4\tanh\frac{\beta}{2}\hbar\omega}\right),\label{eq:specific result}
\end{align}
while in this case we can verify that the right-hand side of the Eq.
\eqref{eq:the JE} $F(\lambda_{t_{f}})-F(\lambda_{t_{0}})=0$. Thus
the JE fails but we observe that in high-temperature limit the factor
in the right-hand side of Eq. \eqref{eq:specific result} is

\begin{align}
\frac{1}{2}-\frac{\beta\hbar\omega}{4\tanh\frac{\beta}{2}\hbar\omega}=- & \frac{\left(\beta\hbar\omega\right)^{2}}{24}+O(\left(\beta\hbar\omega\right)^{4})
\end{align}
and the JE is recovered. We know that the quantum framework we used
in this typical model is not consistent in the sense of Griffiths
\cite{Griffiths1984}, though it is relevant to the single molecule
experiment. For a quantum framework to be consistent the consistency
conditions need to be satisfied, explicitly which means $\mathrm{Tr}[\rho^{ini}K^{\dagger}(Y)K(Y')]=0$
for all the different quantum history $Y$ and $Y'$. As Griffiths
mentioned, interactions with the environment can sometimes ensure
the consistency of a family of histories which would be inconsistent
were the system isolated. And we noticed that recently Subasi and
Hu have used the influence functional technique to explore quantum
work distribution under the environment-induced decoherent histories
\cite{Subasi2012}. Therefore another value of the work done defined
on the continuous quantum histories may lie in the study of open systems.
In the following, we will briefly explain how to study the quantum
work and heat transfer simultaneously in the framework of continuous
quantum histories for open systems.

We use a simple model to illustrate the main idea, in which the center-shifted
one-dimensional quantum harmonic oscillator is linearly connected
to a Rubin bath \cite{Rubin1971} so that the total Hamiltonian is
\begin{align}
\hat{H}_{tot}(t)= & \hat{H}(t)+\hat{H}_{SB}+\hat{H}_{B},
\end{align}
with $\hat{H}_{SB}=-k_{1}\hat{x}\hat{x}_{1}$ and $\hat{H}_{B}=\frac{1}{2m_{0}}p_{B}^{T}p_{B}+\frac{1}{2}x_{B}^{T}Kx_{B}$.
Where $K$ is the semi-infinite tridiagonal spring constant matrix
consisting of $2k_{1}+k_{0}$ along the diagonal and $k_{0}$ along
the two off diagonals, $p_{B}$ and $x_{B}$ are both semi-infinite
column vectors with elements $\left[p_{B}\right]_{i}=\hat{p}_{i}$
and $\left[x_{B}\right]_{i}=\hat{x}_{i}\: i=1,2,\ldots$ , respectively.
The power operator for the heat transfer is simply the heat current
operator. In this example the heat current operator in the Schr\"{o}dinger
picture describing the heat transfer out of the bath is $\hat{I}=-\frac{k_{1}}{m_{0}}\hat{p}_{1}\hat{x}$,
while the power operator for the work remains the same as before,
i.e., $\hat{A}\left(t\right)=-\dot{\lambda}\left(t\right)m\omega^{2}\left[\hat{x}-\lambda\left(t\right)\right]$.
The common eigenstates of $\hat{I}$ and $\hat{A}\left(t\right)$
are easily found to be $\left|x\right\rangle \otimes\left|p_{1}\right\rangle $,
which can be used to construct the projectors so that the continuous
quantum histories can be similarly established as before. Thus generally
one can study the work done and heat transfer in the same set of quantum
histories and in this case the time evolution is due to the total
Hamiltonian $\hat{H}_{tot}(t).$


$Summary$\rule[0.5mm]{0.05\columnwidth}{1pt}We consider the general definition of work distribution under continuous
quantum histories. And we obtain the exact quantum work distribution
for the model a center-shifted one-dimensional quantum harmonic oscillator
in the sense of strong measurement, which turns out to be Gaussian
profile opposite to the prediction of Chernyak mentioned in the introduction
part. The key point is that we carefully introduce Wick's rotation
for the transition probability. For closed systems, the failure
of the quantum version of JE under this definition of work is observed.
Furthermore, the extension to open systems is briefly explored and
it has been noticed that generally the work done and heat transfer
can be calculated in the same set of quantum histories.
\begin{acknowledgments}
We would like to thank Prof. Tapio Ala-Nissilä, Yuan Luo, Bijay Kumar
Agarwalla and Sha Liu for insightful discussions.
\end{acknowledgments}

\end{document}